\documentclass[reprint,article,showpacs,preprintnumbers,amsmath,amssymb,citeautoscript,aip,superscriptaddress,floatfix]{revtex4-1}
\usepackage{graphicx,wasysym}
\usepackage{dcolumn}
\usepackage{hyperref}
\usepackage{color}
\usepackage{bm,dsfont,amsfonts}
\usepackage{mathtools}
\usepackage{braket}
\usepackage[normalem]{ulem}

\newcommand{\A}{\mathcal{A}}

\newcommand{\J}{\mathcal{J}}
\renewcommand{\L}{\mathcal{L}}
\newcommand{\Id}{\mathds{1}}

\renewcommand{\P}{\mathcal{P}}

\newcommand{\Ex}{\mathbb{E}}

\DeclareMathOperator{\Tr}{Tr}

\setcitestyle{super}

\newcolumntype{P}[1]{>{\centering\arraybackslash}p{#1}}

\makeatletter

\begin{document}
\author{Michelle C. Anderson}
\affiliation{%
Department of Chemistry, University of California, Berkeley 
}
\author{Amro Dodin}
\affiliation{%
Department of Chemistry, University of California, Berkeley 
}
\affiliation{%
Chemical Sciences Division, Lawrence Berkeley National Laboratory
}
\author{Thomas P. Fay}
\affiliation{%
Department of Chemistry, University of California, Berkeley 
}
\author{David T. Limmer} \email{dlimmer@berkeley.edu}
\affiliation{%
Department of Chemistry, University of California, Berkeley 
}
\affiliation{%
Kavli Energy NanoSciences Institute, University of California, Berkeley 
}\affiliation{%
Chemical Sciences Division, Lawrence Berkeley National Laboratory
}
\affiliation{%
Materials Sciences Division, Lawrence Berkeley National Laboratory
}
\preprint{1}

\begin{abstract}

We introduce a general definition of a quantum committor in order to clarify reaction mechanisms and facilitate control in processes where coherent effects are important. With a quantum committor, we generalize the notion of a transition state to quantum superpositions and quantify the effect of interference on the progress of the reaction. The formalism is applicable to any linear quantum master equation supporting metastability for which absorbing boundary conditions designating the reactant and product states can be applied. We use this formalism to determine the dependence of the quantum transition state on coherences in a polaritonic system and optimize the initialization state of a conical intersection model to control reactive outcomes, achieving yields of the desired state approaching 100\%. In addition to providing a practical tool, the quantum committor provides a conceptual framework for understanding reactions in cases when classical intuitions fail. 
\end{abstract}

\title{
Coherent control from quantum committment probabilities
}

\maketitle

\section*{Introduction}
Understanding mechanisms of reactions evolving quantum mechanically is difficult, as delocalization, tunneling and interference preclude equating a reaction with the classical idea of motion over a barrier. The committor, which is the probability of the system to complete a reaction,\cite{onsager1938initial,hummer2004transition}  defines an ideal reaction coordinate, and is used to identify transition states and mechanisms in classical reactions. Previous work has recently extended the notion of the committor to systems in which transition states are delocalized and the dynamics are non-adiabatic,\cite{anderson_schile_limmer,schile_limmer} but has stopped short of dealing with the dynamical consequences of coherences, due to the complications that arise in defining trajectories in such cases.\cite{manzano2015nonequilibrium} Here we overcome these previous limitations by introducing a definition of a quantum committor consistent with any quantum dynamical map in which the propagator can be formed as a linear superoperator acting on the system density matrix with certain states designated as absorbing boundary conditions.
This allows us to explore how mechanisms and transition states are influenced by interference, in addition to quantum uncertainty. We have applied this framework to investigate how the transition state in a thermal barrier crossing reaction in a polaritonic system is affected by coherences and to coherently control the outcome of vertical relaxation of a system with a conical intersection. 

Defining a quantum committor, or splitting probability, requires the existence of multiple metastable states. The meaning of metastability in quantum systems remains a topic of active investigation\cite{patra_2015,deleon_1989}. In cases where a system is in contact with a decohering environment, quantum metastability can be associated with a spectral gap in the generator of the dynamics, evident in a separation of relaxation timescales\cite{katarzyna_2016,katarzyna_2021,brown_2023}.
Imposing definitions of reactants and products on the system, quantum incarnations of transition path sampling have been applied using unravelled quantum master equations,\cite{schile_limmer} and surface hopping models.\cite{reiner_2023,coffman_2023} Ensembles of quantum trajectories have been analyzed through extensions of transition path theory,\cite{anderson_schile_limmer} however such analysis requires the dynamics to be secular.\cite{breuer_petruccione,schile_limmer} This assumption decouples coherences from populations in the energy eigenbasis, mapping the system dynamics to a classical Markov process. Such a mapping eschews any influence of quantum coherences by assigning each energy eigenstate a committor probability. Interference effects significantly impact reaction rates and pathways in many  systems of chemical interest, and the inability to address coherent effects seriously limits the applicability of these quantum committor methods.\cite{keefer_schnappinger,schlosshauer,yang_2020,duan_2016} This is especially true for systems with conical intersections\cite{duan_2016,qi_2017,wang_2019,zhu2016non,worth_cederbaum,schile2019simulating,halpern2020fundamental} or strongly coupled to light\cite{mandal_2019,bhuyan_2023,lindoy_nature,fiechter_2023,li_nitzan_subotnik} for which the secular approximation breaks down due to gaps in the system eigenspectrum being small compared to the rate of relaxation of the bath.  

To investigate the effects of coherences on reactive behavior, we have defined a quantum mechanical committor that can be assigned to any quantum state or superposition. This idea is developed in the context of a partially secularized Redfield\cite{redfield,trushechkin} master equation, but is general provided an accurate 
 Markovian description of thermalization and decoherence. With this generalization of the  committor, we have characterized coherent quantum effects on the transition state in a polaritonic system, finding profound alterations in strong light-matter coupling cases. We have further employed knowledge of the committor to engineer ideal initial states in which interference effects guarantee relaxation into a preferred product in a conical intersection model. The ease of selecting these ideal initial conditions opens intriguing possibilities. Given recent advances in laser technology which allow monitoring and manipulation of systems on timescales relevant to electronic dynamics,\cite{zhu_1997,kraus_2009,arasaki_2011} the ability to determine  initial states for desired reactive outcomes has potential applications in coherent quantum control where biasing photoisomerization results or controlling relaxation pathways is a common goal. \cite{lucas_2014,vogt_2005,herek_2002}

\section*{Partial Secular Master Equations}
To preserve the influence of coherences while also guaranteeing complete positivity of the density matrix, we employ a partial secularization of the Redfield master equation. The Redfield master equation\cite{redfield,nitzan} describes the evolution of a quantum system with total Hamiltonian $H=H_S + H_B + H_I$, where $H_S$ is the Hamiltonian of the system, $H_B$ is the Hamiltonian of the thermal bath, and $H_I= \sum_k S_k \otimes G_k$ represents the weak, bilinear coupling between the system and bath. Operator $S_k$ is an operator on the system and $G_k$ is the $k$'th statistically independent operator on the bath. 

The Redfield master equation describing the evolution of system density matrix, $\rho(t)$, where the bath degrees of freedom have been traced out, can be formulated as\cite{breuer_petruccione} 
\begin{eqnarray}
    \dot{\rho}(t) &=& -i \left[ H_S + {H}_{LS}, {\rho}(t) \right] + \sum_{\omega,\omega', k}  \gamma_{k}(\omega',\omega)  e^{i (\omega' - \omega)t}\notag \\ 
    &&\left(S_{k} (\omega) {\rho}(t) S^*_{k} (\omega') - \frac{1}{2} \{ S^*_{k} (\omega) S_{k} (\omega),\; {\rho}(t) \} \right) \label{Eq1}
\end{eqnarray}
where $\hbar=1$ and $\dot{\rho}(t)$ indicates the time derivative of $\rho(t)$. The operator $S_k(\omega)$ is the portion of the operator $S_k$ which produces density transfer between any eigenstates $j$ and $i$ such that $\omega=\epsilon_i - \epsilon_j$ are the Bohr frequencies of the system Hamiltonian, with $\epsilon_i$ being the $i$'th energy eigenvalue with eigenvector $|i\rangle$. Specifically, 
\begin{equation}
S_k(\omega)=\sum_{\epsilon_i - \epsilon_j = \omega } |i\rangle \langle i| S_k |j \rangle \langle j|,
\end{equation}
is the operator producing density transfer between eigenstates separated by an energy gap of $\omega$. We define $S^*_k(\omega)$ as the conjugate transpose of $S_k(\omega)$. The factors $\gamma_k({\omega',\omega})$ are given by
\begin{equation}
\gamma_k({\omega',\omega}) = \Gamma_{k}({\omega}) + \Gamma^*_{k} ({\omega'}),
\end{equation}
and the one sided Fourier transform of the bath correlation function define
\begin{equation}
\Gamma_{k} ({\omega}) = \int_0^{\infty} \mathrm{Tr_B} [ \tilde{G}^*_k (s) \tilde{G}_k (0) \sigma_B  ] e^{i {\omega} s} \; ds,
\end{equation}
with $\sigma_B$ being the density matrix of the thermal equilibrium state of the bath. The interaction picture bath operators, $\tilde{G}_k$, are $\tilde{G} = e^{i H_B t}G e^{-i H_B t}$. The Lamb shift Hamiltonian,
\begin{equation}
{H}_{LS} = \sum_{k \omega \omega'} \Pi_{k}( \omega', \omega) S^*_{k} (\omega')S_{k} (\omega),
\end{equation}
includes factors,
\begin{equation}
\Pi_k({\omega'},\omega) = \frac{1}{2i} \left[ \Gamma_{k} ({\omega}) - \Gamma^*_{k} ({\omega'}) \right],
\end{equation}
which are also built from one-sided Fourier transforms. This expression for the evolution of the reduced density matrix assumes only weak system-bath coupling and a Markovian bath.\cite{levy2021response} 

Due to the cross terms generated by operators $S_k(\omega)$ and $S_k(\omega')$ in the summation, the terms which preserve the influence of coherences on the evolution, this equation cannot be written in the form of a dynamical semigroup. The commonly applied secular Redfield master equation\cite{breuer_petruccione,nitzan,balzer_stock} achieves dynamical semigroup form by arguing that, in the weak coupling  limit, the unitary oscillation of the coherent phase is much faster than population transfer. Thus the complex exponential under the summation in Eq.~\ref{Eq1}, will oscillate many times before significant population is transferred,  averaging to zero unless $\omega' - \omega = 0$. The nonsecular terms, any which do not fulfill this requirement, can be neglected, decoupling populations from coherences and yielding the fully secular quantum master equation which is in the form of a dynamical semigroup and thus guarantees positivity of $\rho$.\cite{breuer_petruccione} Any coherences in the initial density matrix at time zero decay exponentially with no influence on the populations. 
When a near degeneracy occurs in the system eigenspectrum, the assumption that we may neglect the average influence of the quickly oscillating exponential term no longer holds and there is significant influence of quantum coherences on the evolution of populations.\cite{egorova_2001,cattaneo_2019} 

An alternative partial secularization which preserves important coherences\cite{trushechkin,cattaneo_2019} can be derived under the piecewise flat secular approximation. This approximation demands that the spectral density describing the bath does not change over energy ranges encompassing all eigenstates in which coherent effects are relevant. If the coherence between eigenstates $i$ and $j$ is important, then for any $\omega = \epsilon_i - \epsilon_l \; l\ne i,j$ and $\omega' = \epsilon_j - \epsilon_l\; l\ne \; i, j$ the Fourier transforms of the bath correlation function are equal, $\Gamma_k(\omega)=\Gamma_k(\omega')$. Because coherent effects are generally large when energy differences between states are very small relative to system timescales, this approximation is usually not demanding. To apply the approximation, the energy eigenspectrum of $H_S$ is organized into nearly degenerate nonsecular blocks by splitting the Hamiltonian into two parts $H_S = H_S^{(0)} + \delta H_S$, where $H_S^{(0)}$ collapses all nearly degenerate states to the same energy $\bar{\epsilon_i}$ and $\delta H_S$ breaks the degeneracy to recover the initial Hamiltonian.
If $||\delta H_S||\ll ||H_I||$, then we may perform the secular approximation on $H_S^{(0)}$ that retains coherences between degenerate states (i.e. states in the same nonsecular block) and then add in the degeneracy breaking as a perturbative correction. The average energy of the eigenstates in a nonsecular block is designated $\bar{\epsilon_i}$ for nonsecular block $i$. The energy difference between each block is denoted $\bar{\omega}_{ij} = \bar{\epsilon_i} - \bar{\epsilon_j}$. A block is assigned a first eigenstate entry and if the next higher energy eigenstate is less than a limit, $\hbar \gamma$, higher in energy, it is added to the block. The difference in energies between the newly added eigenstate and the next higher eigenstate is tested and eigenstates added until one exceeds $\hbar \gamma$. This highier eigenstate is the first entry in the next block. The factor, $\gamma$, should formally be equal to the total rate of population exit from the eigenstate under consideration.\cite{trushechkin} 

The operators, $S_k(\bar\omega),$ in the partial secular quantum master equation are formed from the system-bath interaction Hamiltonian such that each operator is either the dephasing operator with $\bar \omega=0$ or defines a jump process between nonsecular blocks $i$ and $j$, not a single pair of eigenstates, such that 
\begin{equation}
S_k(\bar{\omega}) = \sum_{\bar{\epsilon}_{BL(i)} - \bar{\epsilon}_{BL(j)} = \bar{\omega}} S_k(\omega)
\end{equation}
 where $\bar{\epsilon}_{BL(i)}$ refers to the average energy of the nonsecular block to which eigenstate $i$ belongs.
After defining nonsecular blocks and relevant jump operators, the partial secular master equation is identical to the secular equation save that all $\omega$ are replaced by $\bar \omega$. This reduces Eq.~\ref{Eq1} to 
\begin{eqnarray}
\label{eq:partial_secular}
\dot{\rho}(t) &=& -i \left[H_S + H_{LS} \right] + \sum_{\bar{\omega},k} \gamma_{k}(\bar\omega,\bar\omega) \\ &&\left(S_k (\bar\omega) \rho(t) S_k^* (\bar\omega) - \frac{1}{2} \{ S_k^* (\bar\omega) S_k (\bar\omega), \rho(t) \}\right), \notag
\end{eqnarray}
which is trace preserving and completely positive. The Lamb shift Hamiltonian becomes block diagonal, its structure corresponding with that of the selected nonsecular blocks. Coherent evolution occurs within the nonsecular blocks, facilitated both by the Lambshift Hamiltonian and the action of $S_k(\bar\omega)$ operators under the summation, which produce cross terms linking coherences and populations within the nonsecular blocks only. Coherences between eigenstates in different nonsecular blocks still decay exponentially without influencing population dynamics, but those within the blocks can increase or decrease under the influence of these partially secularized operators and may have significant impacts on population dynamics. In the case that each block contains a single eigenstate, meaning all coherences are decoupled from all populations, the partial secular equation reduces to the full secular approximation.

\section*{Coherent quantum committor}
The committor is conventionally defined as the probability, given sets of system states $A$ and $B$, that the system will visit a state in $B$ before a state in $A$, denoted $P_{B|A}$.  In the context of chemical reactions set $A$ can be considered the reactants and $B$ the products, both of which are expected to be metastable as defined by the separation of timescales evident in the eigenspectrum of the propagator.\cite{katarzyna_2016,brown_2023} In the context of the partial secular Redfield dynamics we consider, this requires that no state of $A$ or $B$ be in a nonsecular block with any state not also in $A$ or $B$, such that the projective measurement implied by the conditioning in the reactive trajectory ensemble\cite{schile_limmer} does not destroy coherences relevant to the reactive dynamics. 

Under these assumptions, the committor can be evaluated by constructing a superoperator in the energy eigenbasis such that $\dot{\rho}(t) = \mathcal{L} \rho(t)$, modified to enforce absorbing boundary conditions in all states within the sets of states in $A$ and $B$.\cite{breuer_petruccione,sharpe2021nearly} For our partial secular equation, absorbing boundary conditions are imposed by changing the relevant rates such that $\gamma_k (\bar{\omega}) = 0 \; \forall \; \bar{\omega} = \bar{\epsilon_j} - \bar{\epsilon_i} \; \; j \in A \cup B$. If this approach were applied to a master equation in which coherences linked $A$ and $B$ to the rest of the system, these would have to be approximated as zero, which would not always be a justifiable assumption. The modified $\mathcal{L}$ with absorbing states is denoted by $\mathcal{L}'$. The time integrated flux into any eigenstate $b \in B$ is then
\begin{equation}
\label{eq:committor_component}
V_{b} (\rho_{ij}) = \left( \lim _{t\rightarrow \infty} e^{\mathcal{L'}t} \rho_{ij} \right)_{b,b}
\end{equation}
where $\rho_{ij} = |i \rangle \langle j|$ designates an  initial condition and the subscript $b,b$ indicates that the $b$'th population of the density matrix has been extracted after propagation. The time integrated flux for the entirety of set of states in $B$ is given by the sum $V_{B|A}(\rho_{ij})=\sum_{b \in B} V_b(\rho_{ij})$. Together, $\rho_{ij}$ for all $i$ and $j$ define a basis for any initial density matrix. By calculating $V_{B|A}$ for all basis entries, we can easily calculate $V_{B|A}$ for any density matrix $\rho = \sum_{ij} c_{ij} \rho_{ij}$ by summing over the contributions of each basis state. Because coherences which are not contained in a nonsecular block cannot have any impact on population dynamics, $V_b$ only needs to be calculated for the subset of the basis where it is not zero by definition. The final summation,
\begin{equation}
\label{eq:committor}
P_{B|A}(\rho) = \sum_{i,j} c_{ij} V_{B|A} ( \rho_{ij} ),
\end{equation}
is the probability that a system initialized in state $\rho$ will reach state $B$ before any other state in $A$. 

\section*{Transition States under vibrational strong coupling}
In our first application of the general quantum committor method, we investigated a polariton model similar to that used by Lindoy and coworkers\cite{lindoy_nature}. Vibrational polaritons form when the vacuum photon frequency of a microcavity is near resonance with the frequency of a vibrational mode of a molecule in the cavity. Experimental investigations have indicated that resonance effects in cavity confined polariton systems can influence ground state reactivity, inverting the preferred formation of products in bond breaking reactions.\cite{thomas_lethuillier-karl,thomas_george_shalabney,lather_bhatt_thomas} Theoretical work has studied potential origins for the resonance effects both by classical\cite{li_mandal,philbin_2022,galego_2019,li_mandal_2021} and quantum mechanical\cite{lindoy_nature,schafer_2022,pino_feist,yang_cao} means, with evidence indicating that a fully quantum mechanical approach to the dynamics is likely necessary to understand this phenomenon\cite{lindoy_nature,li_nitzan_subotnik,fiechter_2023,anderson2023mechanism}. 

We have studied the quantum transition state in a polariton model as a function of light-matter coupling using the generalized quantum committor. This model includes a single photon and proton degree of freedom under a Shin-Metiu\cite{shin_metiu} model employing the Pauli-Fierz\cite{pauli_fierz} Hamiltonian which treats light and matter coordinates on the same quantum footing. The Hamiltonian is given by 
\begin{equation}
H_S = \frac{P^2}{2M} + U(R) +\frac{p_c^2}{2} + \frac{\omega_c^2}{2} \left(q_c + \sqrt { \frac{2}{\omega_c}} \eta_c \mu(R) \right)^2,
\end{equation}
where $P$ is the momentum operator of the proton coordinate with position $R$, $M$ is the proton mass, $U(R)$ describes a quartic proton potential energy surface, $q_c$ is the position operator for the photon, $p_c$ is the momentum operator for the photon, $\omega_c$ is the photon frequency, $\mu(R)$ is the proton dipole operator, and $\eta_c$ is the light-matter coupling strength.
The system-bath coupling is bilinear and independent for both the proton and cavity modes
\begin{equation}
H_I = R \otimes \sum_k c_{k,1} r_{k,1} + q_c \otimes \sum_k c_{k,2} r_{k,2},
\end{equation}
and the spectral density describing the coupling strengths, $c_{k,\alpha}$, for each bath $\alpha=\{1,2\}$ is,
\begin{equation}
    J_{\alpha}(\omega) = \sum_k {c_{k,\alpha}^2} \delta(\omega -\omega_{k,\alpha}) =\eta \omega e^{-|\omega|/\omega_b},
\end{equation}
an Ohmic exponential form, where $\omega_b$ is the cutoff frequency and $\eta$ is the coupling strength for both baths. Appendix A compiles the parameters used for this study.

\begin{figure}[t]
\begin{center}
\includegraphics[width=8.5cm]{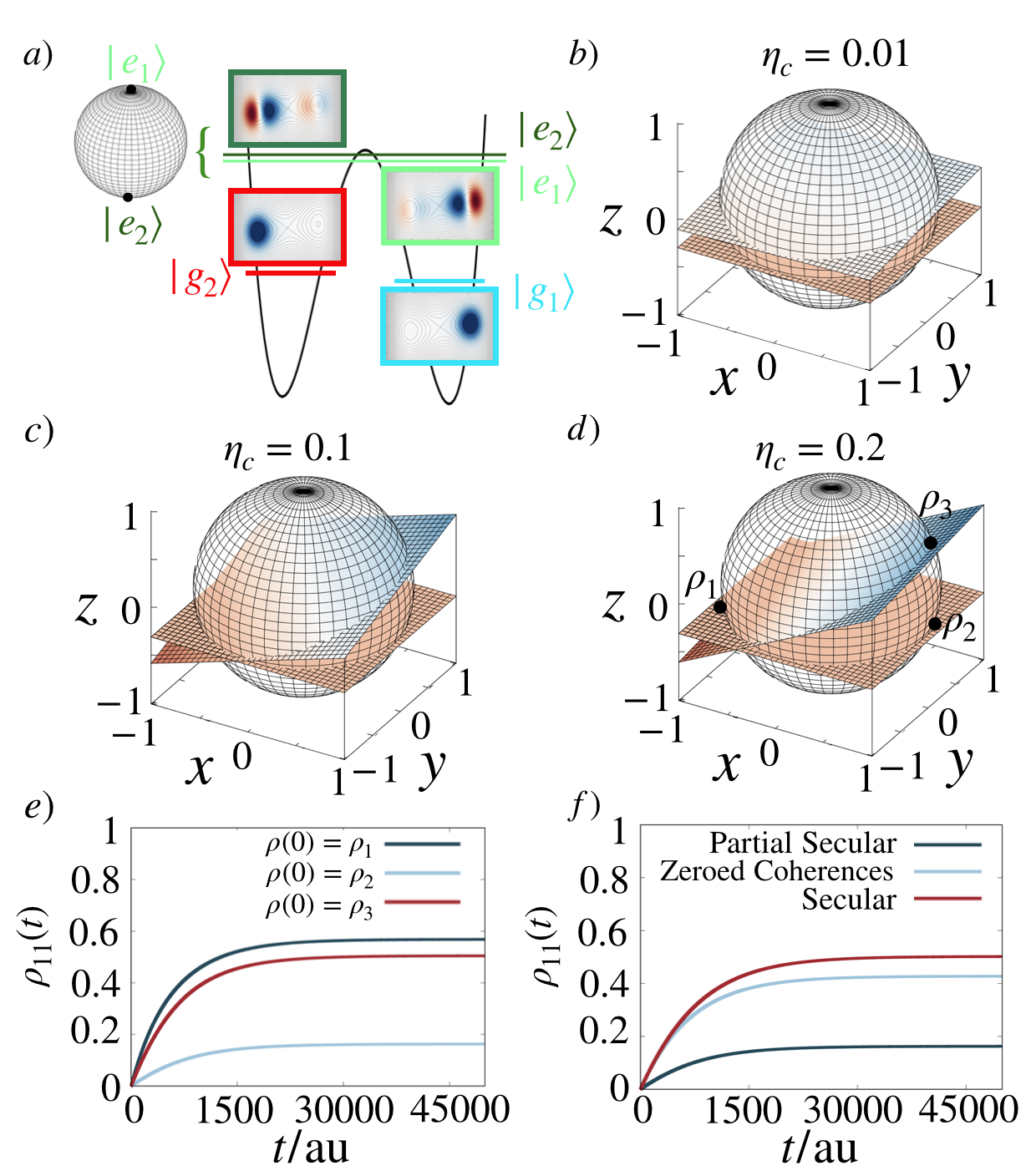}
\caption{a) The potential energy surface, $U(R)$ and lowest four  eigenstates plotted on the adabiatic potential energy surface beside an illustration of the $|e_i\rangle$ states on the Bloch sphere.  The partial secular (upper) and fully secular (lower) committor planes through the Bloch sphere for b) weak $\eta_c=0.01$, c) intermediate $\eta_c=0.1$, and d) strong $\eta_c=0.2$ light-matter coupling. e) The population of state $|g_1\rangle$ as a function of time given initialization in $\rho(0)$ equal to each of the three marked density states in panel d). f) Population in $|g_1\rangle$ given initialization in $\rho(0)$ equal to $\rho_2$ as indicated in panel d) for master equations that are partial secular, secular, and partial secular with the coherences manually set to zero after each time step.}
\label{fig1}
\end{center} 
\end{figure}

We treat the four lowest energy eigenstates only, with the first and second eigenstates identified as $|g_1\rangle$ the reactant $A$ state and $|g_2 \rangle$ the product $B$ state. The third and fourth eigenstates, $|e_1 \rangle$ and $|e_2 \rangle$, form a nonsecular block. The system setup, including the four relevant eigenstates, is shown in Fig. \ref{fig1} a). The excited state set spanned by $|e_1\rangle$ and $e_2\rangle$ can be represented by a Bloch sphere with the eigenstates at its poles. We define the committor surface as the locations in the Bloch sphere where the probabilities to arrive in state $|g_2\rangle$ or $|g_1\rangle$ are both $0.5$, a generalized transition state for quantum systems. Due to the linear nature of the propagator, the committor surface is guaranteed to be a plane. The committor plane is easily found by determining $V_{ g_1   | g_2 }(\rho_{ij})$ for a basis of the density matrix in the nonsecular block, solving for conditions at which $P_{g_1   | g_2 }(\rho)=0.5$ and transferring the obtained solution into Bloch sphere coordinates. To inspect the influence of quantum coherent effects, the committor plane can be defined for the fully secular master equation as well, in which case the plane must be defined by a single $z$ coordinate.

We solved for the committor planes for the partially and fully secular master equation for three different values of light-matter coupling strength. In the weak light-matter coupling case, Fig. \ref{fig1} b), the partial secular plane is nearly flat. However, it lies above the fully secular committor plane. This is due to the piecewise flat spectral density approximation in the partial secular master equation, which alters individual jump rates from those found for equivalent jumps in the fully secular master equation. As the light-matter coupling increases, the fully secular committor plane remains largely unchanged whereas the partial secular committor plane acquires significant tilt, indicating the increasing importance of quantum coherences. This is evident for moderate coupling in Fig. \ref{fig1} c) but especially obvious for the strong coupling case in Fig. \ref{fig1} d). The angle along the $x$ direction indicates a significant influence exerted by the real part of the coherence between $|e_1\rangle$ and $|e_2\rangle$ on determining the outcome of the reaction. This implies that the transition state is represented by a particular set of coherent states at the top of the barrier.  

To demonstrate the influence of coherences on dynamics in the strongest coupling case, three points on the surface of Bloch sphere, two on the fully secular committor plane and one on the partial secular committor plane, were selected as initial conditions to run partial secular dynamics, displayed in Fig. \ref{fig1} e). Initial conditions designated by $\rho_1$ and $\rho_2$ have identical $z$ values, meaning identical populations in $|e_1\rangle$ and $|e_2\rangle$. However, the initial state coherences are different and initialization in $\rho_2$ results in a final population in $|g_1\rangle$ below $0.5$ whereas initialization in $\rho_1$ results in a final population of $|g_1\rangle$ above $0.5$, demonstrating the strong influence of quantum coherent effects on dynamics. By contrast the initial condition $\rho_3$ begins on the partial secular committor plane and thus reaches a final state with population $0.5$, representing the equal likelihood of evolving to the product state or back to the reactant state. The particular importance of the coherence on the population dynamics is confirmed in Fig. \ref{fig1} f), in which we ran partial secular, fully secular, and zeroed coherence dynamics starting from initial $\rho_2$. Zeroed coherence propagation indicates that the partial secular master equation was employed but coherences in the system were manually set to zero at each timestep, removing their influence. This process artificially removes the population-coherence coupling while making the piecewise flat spectral density approximation. Without effects from coherences, the relaxation dynamics are nearly identical to the fully secular relaxation dynamics.

\section*{Coherent Control of a Conical Intersection}
Conical intersections are commonplace in large molecules where they offer ultrafast, nonradiative relaxation pathways,\cite{neville,matsika,curchod2018ab} making them key in isomerization of molecular photoswitches,\cite{muzdalo_2018,bull_2018} light harvesting mechanisms in photosynthetic organisms,\cite{song_nam,savolainen_2008} and damage mitigation in DNA. \cite{barbatti_aquino} At a conical intersection, the splitting between states is expected to become small, enhancing the role of coherent and interference effects in relaxation following photoexcitation. With the ability to probabilistically understand reaction outcomes with the committor, we show that this knowledge can be translated into a control strategy by preparing specific states to optimize the relaxation outcome following photoexcitation. 

The model we consider is similar to a minimal pyrazine model, with parameters modified to produce two metastable wells in the excited state.\cite{chen_lipeng,schile2019simulating,duan_2016,duan_2018,qi_2017} The system Hamiltonian is given by,
\begin{equation}
H_S = \sum_{i=1,2} | \phi_i \rangle h_i \langle \phi_i | + \left(| \phi_1 \rangle \langle \phi_2 | + | \phi_2 \rangle \langle \phi_1 | \right) \lambda Q_c,
\end{equation}
with $i$ indexing the diabatic electronic states, $|\phi_i\rangle$, and
\begin{equation}
h_i = 1/2 \sum_{j=c,t} \hbar \omega_j \left\{P_j^2 + Q_j^2 \right\} + E_k + \kappa_k Q_t
\end{equation}
where $\lambda$ is the diabatic coupling strength, $Q_{j}$ is the dimensionless coupling ($j=c$) or tuning ($j=t$) coordinate, $P_j$ is the dimensionless momentum for $Q_j$, $\kappa_i$ is the displacement along $Q_t$ for each diabatic electronic state, $\omega_{j}$ is the coupling ($j=c$) or tuning frequency ($j=t$) frequency and $E_i$ is the diabatic electronic state energy. The system-bath coupling Hamiltonian is bilinear in the two modes of the conical intersection and bath modes,
\begin{equation}
H_I= \left( |\phi_1\rangle \langle \phi_1| + |\phi_2 \rangle \langle \phi_2 | \right)\sum_\alpha \sum_{j=c,t}  c_{\alpha,j} r_{\alpha,j} Q_j ,
\end{equation}
where $r_{\alpha,j}$ are dimensionless bath oscillator coordinates, and diagonal in the diabatic electronic states. The spectral density again takes Ohmic form,
\begin{equation}
  J_j(\omega) = \eta e^{-|\omega|/\omega_b} \;\quad j = c,t,
\end{equation}
where $\omega_{b}$ is the cutoff frequency and $\eta$ is the system-bath coupling strength. The parameters of the model employed are summarized in Appendix A.

\begin{figure}[t]
\begin{center}
\includegraphics[width=8.6cm]{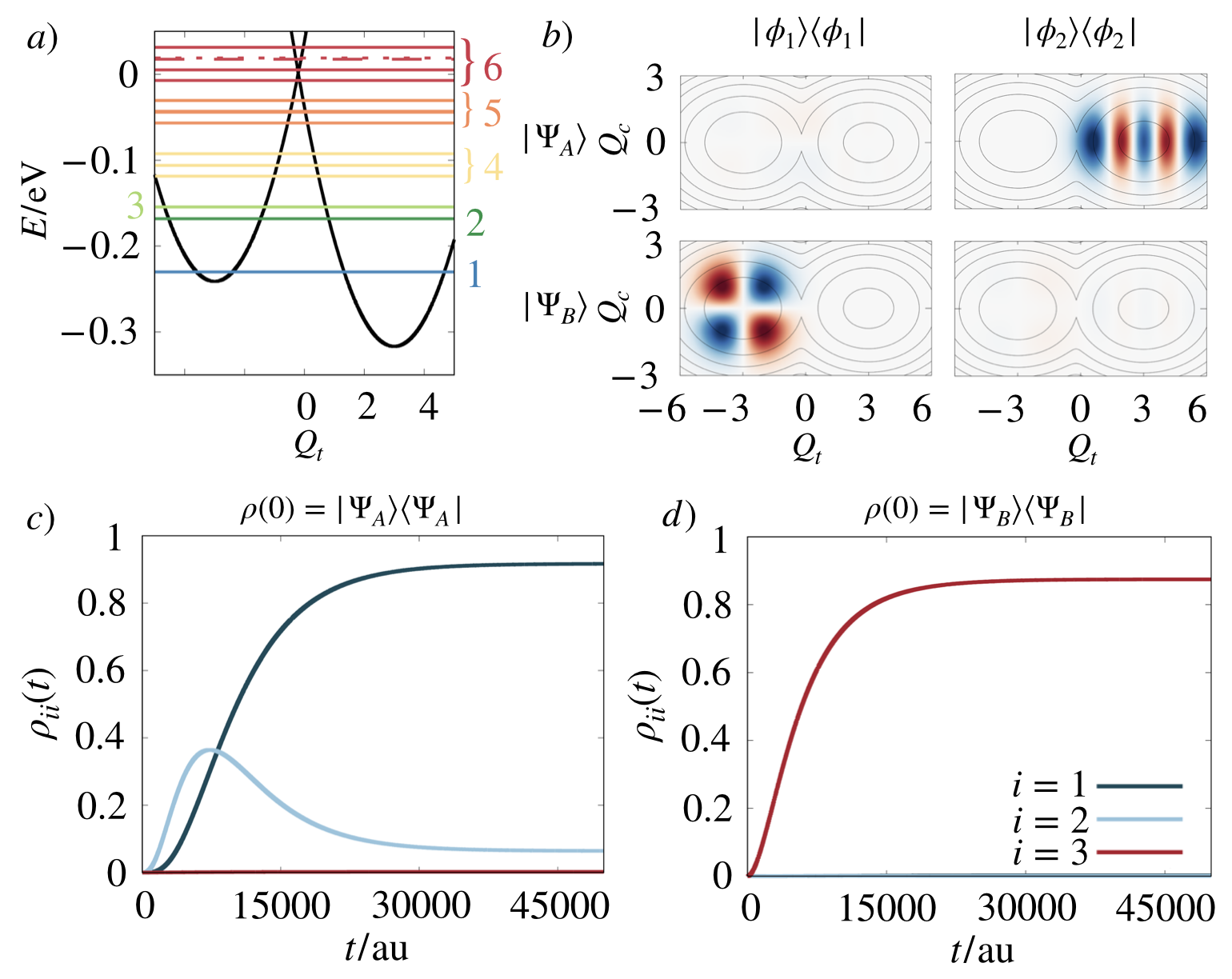}
\caption{a) A cut through the diabatic electronic potential energy surfaces of the conical intersection (black) at $Q_c=0$ and the energies of the fifteen eigenstates of the conical intersection, color coded by nonsecular block, with eigenstates 13 and 14 marked with dashed and dotted lines respectively. b) The real parts in both diabatic electronic states, $|\phi_1 \rangle$ and $|\phi_2\rangle$, of the optimal initial wavefunction for arrival in eigenstate 1, $|\Psi_A\rangle$, and the optimial initial wavefunction for arrival in eigenstate 3, $|\Psi_B \rangle$. c) Populations in the lowest three eigenstates as a function of time following initialization of the system in $\rho(0) = |\Psi_A \rangle \langle \Psi_A|$. d) Populations in the lowest three eigenstates as a function of time following initialization of the system such that $\rho(0) = |\Psi_B \rangle \langle \Psi_B|$.}
\label{fig3}
\end{center} 
\end{figure}

An illustration of the structure of the system, including individual eigenstates, non-secular blocks, and diabatic potential energy surfaces is shown in Fig. \ref{fig3} a). The lowest two eigenstates are localized in a state at positive $Q_t$, and we group them into the reactant state $A$, while the third lowest eigenstate is localized at negative $Q_t$ and considered the product state, $B$. Each of these states is its own nonsecular block. There are four other nonsecular blocks, with eigenstates 13 and 14 in block $6$ at the top of the barrier highlighted with dashes. These states are extremely close in energy, which suggests that quantum coherences between them are likely to be important. 

We have calculated $V_{A|B}$ for the populations and coherences of and between eigenstates 13 and 14. We find that $V_{A|B}(\rho_{14,14})\approx0.85$ and that $V_{A|B}(\rho_{13,13})\approx0.15$, with the time integrated fluxes corresponding well with the diabatic character of these eigenstates. We also find that $V_{A|B}(\rho_{13,14})$ is very large.  To illustrate how understanding the committment probability informs coherent quantum control, we optimized population and coherence to direct evolution of the system into either $A$ or $B$ by optimizing the committor value over all possible coherent superpositions of eigenstates $13$ and $14$. When inspecting the optimized wavefunctions displayed in Fig. \ref{fig3} b), the wavefunction optimized to arrive in $A$, $|\Psi_A\rangle$, with $P_{A|B}(\rho = {|\Psi_A\rangle \langle \Psi_A|})$ approaching 1.0, is nearly entirely localized in the right well. Similarly, the wavefunction optimized to arrive in $B$, $|\Psi_B\rangle$, for which the committor $P_{B|A}(\rho={|\Psi_B\rangle \langle \Psi_B|})$ approaches 1.0, is fully localized in the left well. In other words, this system behaves diabatically despite being highly excited. Thus we can optimize the reactive outcomes by using quantum interference to engineer an initial condition. This optimization is also robust to small changes of the model as illustrated in Appendix B.

The differences in population dynamics following excitation to $\rho(0)= |\Psi_A \rangle \langle \Psi_A|$ in Fig. \ref{fig3} c) is strikingly different from dynamics following excitation to $\rho(0)= |\Psi_B \rangle \langle \Psi_B|$ in Fig. \ref{fig3} d), which is not unexpected given that the initial populations are very different, but reflects the selectivity that could not be achieved by incoherent initialization into a combination of eigenstates 13 and 14. Population in $B$, $i=3$, in Fig. \ref{fig3} c) remains at nearly zero for the entire 50000 au period whereas virtually all population is in $B$ over that period in Fig. \ref{fig3} d). Note that in Fig. \ref{fig3} d) some population has become trapped in what is effectively a dark state in nonsecular block 4 and has not relaxed down to eigenstate 3 over the time period displayed, but will do so in the long time limit.

\begin{figure}[t]
\begin{center}
\includegraphics[width=8.5cm]{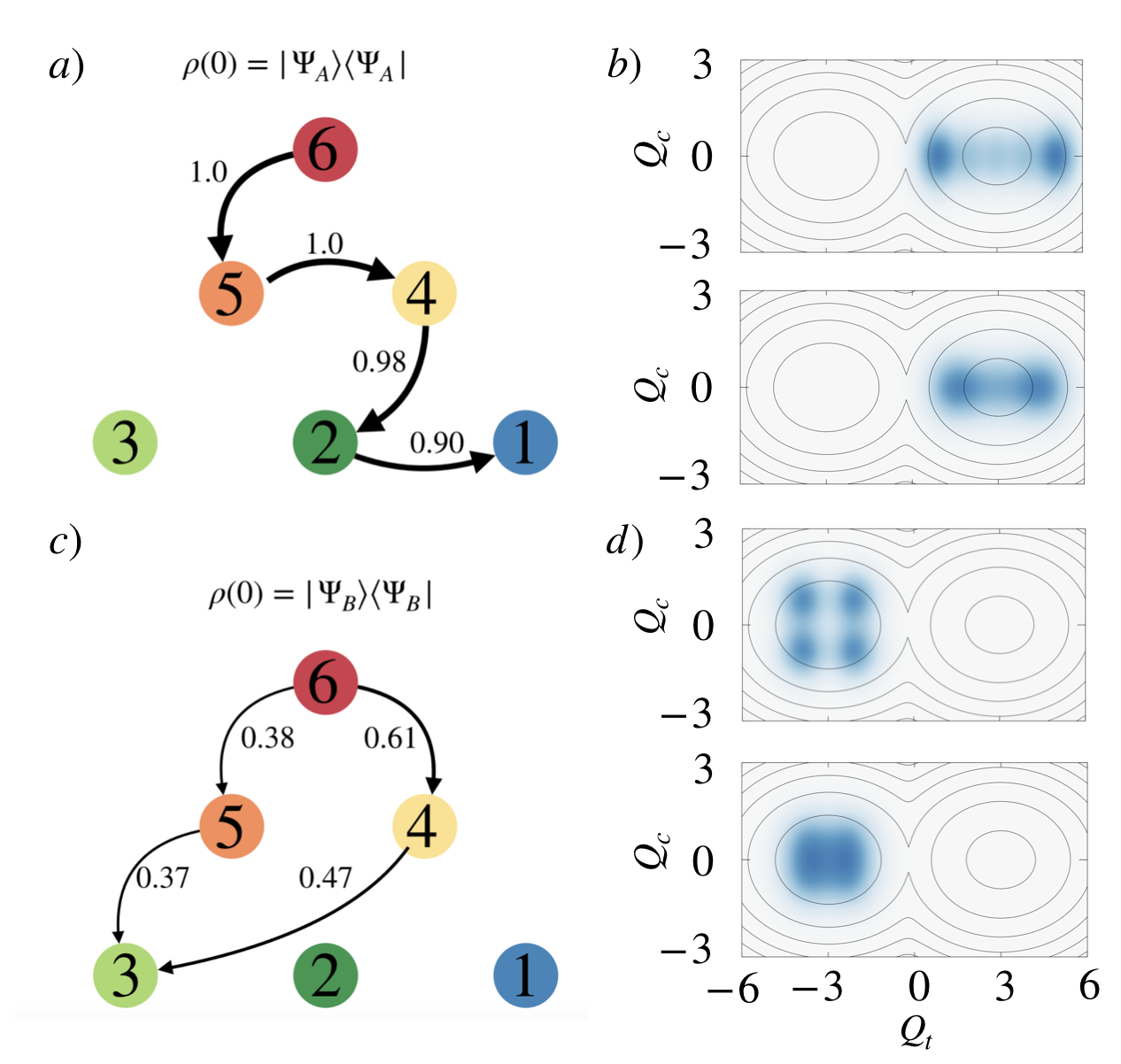}
\caption{a) The net reactive flux traveling between nonsecular blocks for the system initialized in $\rho(0) = |\Psi_A \rangle \langle \Psi_A|$ over a duration of 50000 au. b) Population density in electronic state 2 for the system relaxing from $\rho(0) = |\Psi_A \rangle \langle \Psi_A|$ at $2000 \; \mathrm{au}$ (top) and $5000 \; \mathrm{au}$ (bottom). c) The net reactive flux traveling between nonsecular blocks for the system initialized in $\rho(0) = |\Psi_B \rangle \langle \Psi_B|$ over a duration of 50000 au. Note that not all density arrives in block 3 over this period of time, with a significant amount remaining in block 4. d) Population density in electronic state 1 for the system relaxing from $\rho(0) = |\Psi_B \rangle \langle \Psi_B|$ at $2000 \; \mathrm{au}$ (top) and $5000 \; \mathrm{au}$ (bottom).}
\label{fig4}
\end{center} 
\end{figure}

The location of the density trapped in a dark state can be determined, as can the overall relaxation pathways, by constructing a hidden Markov state model between the nonsecular blocks, with the precise quantum state in the block being the hidden variable which changes over time due to coherent evolution. This hidden state of the block determines when and how density will depart from the block. This removes information about the specific history of individual pathways that would be evident in a quantum jump approach,\cite{anderson_schile_limmer} but is comparatively very simple to implement and provides adequate information for our needs. In Fig. \ref{fig4} a) the relaxation pathway following excitation into $\rho(0) = |\Psi_A \rangle \langle \Psi_A|$ moves between nonsecular blocks with 90\% of density arriving in eigenstate 1 over the simulation period and 8\% remaining in eigenstate 2 with the remaining density largely found in nonsecular block 4. This is a very different pathway than that following excitation to $\rho (0) = |\Psi_B \rangle \langle \Psi_B|$ in Fig. \ref{fig4} c) where two alternate relaxation pathways exist, one following from 6 to 5 to 3 and the other from 6 to 4 to 3, with almost 14\% of density remaining in nonsecular block 4 after 50000 au. Despite the difference in relaxation pathways, inspecting probability density distributions during early times of relaxation in \ref{fig4} b) and d) shows that the system remains localized in both cases, relaxing quickly towards the bottom of the destination well. This method of flux tracing illuminates the general pathways of relaxation, showing that a bifurcation is involved in optimal relaxation to $B$. 

\section*{Conclusions}

By generalizing the definition of the committor to quantum coherent systems, we have quantified the impact of coherences on dynamics in polaritonic systems and conical intersections. Following initial calculations of committors for a basis of the density matrix, the committor value of any quantum state can be trivially obtained. This method can be used to define committors for any linear master equation in which absorbing boundary conditions can be defined, allowing fast determination of initial excitations providing the desired outcome in coherent quantum control problems and access to the committor surface, the quantum generalization of a transition state, which is valuable in the study of quantum reaction mechanisms. 

\section*{Acknowledgements}
This work was supported by the U.S. Department of
Energy, Office of Science, Basic Energy Sciences, CPIMS Program Early Career Research Program under Award DEFOA0002019. D.T.L was supported by the Alfred P. Sloan Foundation.

\subsection*{Appendix A: Simulation Parameters}

Simulations for the polariton were carried out according to parameters in Table \ref{tab2} with a DVR\cite{colbert_miller} basis for $R$ and a harmonic oscillator basis for $q_c$ with respective dimensions of 81 and 60 with $\Delta R$ for the DVR basis of $0.03$ au, although a significantly smaller basis was likely sufficient. The system was then truncated to include the lowest four energy eigenstates. 

The polariton model employs a quartic potential energy surface,\cite{lindoy_nature} 
\begin{equation}
E(R) = \frac{c_{ob}^4}{16 c_{eb}}R^4 - \frac{c_{ob}^2R^2}{2} - c_{cu}R^3,
\end{equation}
with the dipole operator approximated by 
\begin{equation}
\mu(R) = v\mathrm{tanh}(yR) + zR,
\end{equation}
although its exact form does not appear to change the observed trends. The coordinates on the surface of the Bloch sphere used to define density matrix $\rho_1$ in Fig. \ref{fig1} d are $(x,y,z) = (-0.87,-0.17,-0.31)$. Those for $\rho_2$ are $(0.87,0.383,-0.31)$ and those for $\rho_3$ are $(0.87,0.17,0.46)$.  

Simulations for the conical intersection were carried out according to parameters in Table \ref{tab1} with a harmonic oscillator basis for $Q_c$ and $Q_t$ with dimensions 40 and 90 respectively, although a significantly smaller basis was likely sufficient. The system was then truncated to include fifteen eigenstates.

\renewcommand{\arraystretch}{1.5}
\begin{table}
\centering
\begin{tabular}{|p{2.0cm} | p{4.5cm}| }
\hline 
\hline 
Parameter  &  atomic units unless specified  \\
\hline

$\beta$ & 1052.584413  \\
$\omega_b$ & 0.05\\
$\omega_c$ & 0.025344 \\
$\eta$ & 0.2 (unitless)\\
$c_{ob}$ & 0.8  \\
$c_{eb}$ & 0.05  \\
$M$ & 1836  \\
$c_{cu}$ & 0.004  \\
$v$ & -1.7  \\
$y$ & 3.0 (unitless)\\
$z$ & 0.6 \\
\hline
\end{tabular}

\caption{Parameters employed during simulation of the polariton model}. 
\label{tab2}
\end{table}

\renewcommand{\arraystretch}{1.5}
\begin{table}
\centering
\begin{tabular}{|p{2.0cm} | p{4.5cm}| }
\hline 
\hline 
Parameter  &  atomic units unless specified   \\
\hline

$\beta$ & 1052.584413  \\
$\omega_b$ & 0.01\\
$\eta$ & 0.1  (unitless)\\
$E_1$ & -0.00139\\
$E_2$ & 0.00139\\
$\omega_c$ & 0.004116\\
$\omega_t$ & 0.002279\\
$\kappa_1$ & -0.006836\\
$\kappa_2$ & 0.006836\\
$\lambda$ & 0.00091153\\
\hline
\end{tabular}

\caption{Parameters employed during simulation of the conical intersection model. Note that this simulation is carried out with $Q_{c,t}$ and $P_{c,t}$ in dimensionless units.} 
\label{tab1}
\end{table}

\subsection*{Appendix B: Coherent Control Stability Analysis}
\begin{figure}[b]
\begin{center}
\includegraphics[width=8.5cm]{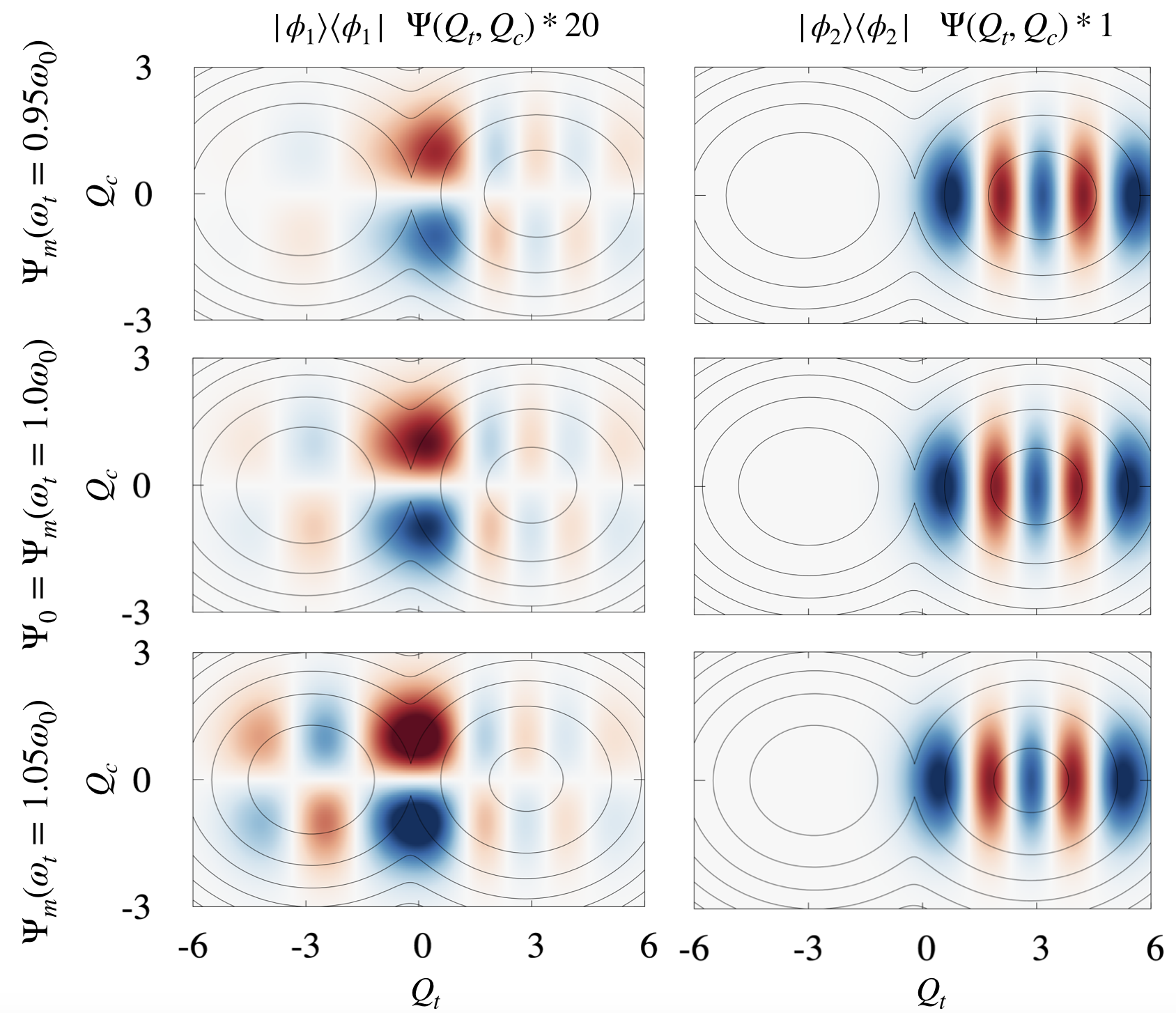}
\caption{The real part of the optimal wavefunctions, $\Psi_m$ (multiplied by a factor for easy viewing) for arrival in eigenstates $1$ and $2$ in the unperturbed and two perturbed systems, where the perturbation adjusts $\omega_t$ relative to the default value of $0.002279\;\mathrm{au}$. The change occurs smoothly. The optimal wavefunction is largely localized in the lower energy well in all cases.}
\label{fig5}
\end{center} 
\end{figure}

In order to assess the sensitivity of the optimized wavefunctions and optimization efficiency to small changes in the Hamiltonian parameters, we determined the optimal initial wavefunction for relaxation into $A$ obtainable from a coherent superposition of eigenstates 13 and 14 as we modified $\omega_t$. The original $\omega_t$ is designated $\omega_0$. Modifications were small to avoid changing the energy ordering of the eigenstates. Optimization of relaxation outcomes was effective for significantly perturbed systems. In Fig. \ref{fig5}, the optimal initial wavefunctions for the case of slightly smaller $\omega_t$, original $\omega_t$ and slightly larger $\omega_t$ are seen to be very similar, although as $\omega_t$ increases the amount of wavefunction density located in the metastable well, diabatic electronic state 1, increases, and, correspondingly, efficiency decreases somewhat.

\begin{figure}[t]
\begin{center}
\includegraphics[width=8.6cm]{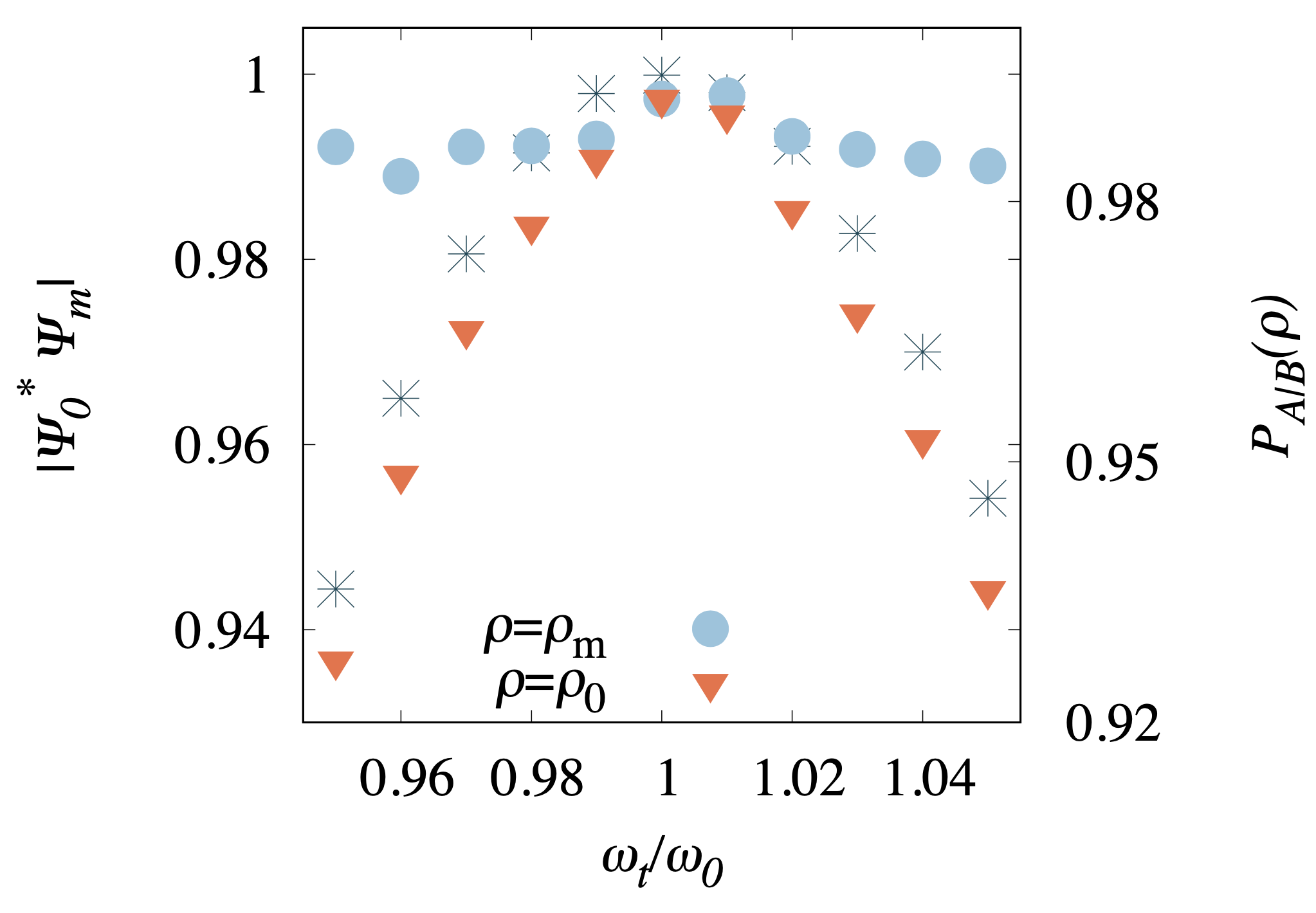}
\caption{Asterisk symbols indicating the overlap between the optimal wavefunction in the unperturbed case and the optimal wavefunction in the perturbed case, $|\Psi_0^* \Psi_m|$, together with committor to $A$, designated $P_{A|B}(\rho)$, as $\omega_t$ is modified for both the optimal initialization density matrix for the unperturbed system, $\rho_0$, and the perturbed optimal initialization density matrix, $\rho_m$.}
\label{fig6}
\end{center} 
\end{figure}

Comparing the overlap between the optimal wavefunction for the unperturbed system with $\omega_t = \omega_0$ and the optimal wavefunction for systems with perturbed $\omega_t$ in Fig. \ref{fig6}, the overlap decreases in tandem with the fall in the committor value obtained from employing the optimal wavefunction for $\omega_t=\omega_0$ as an initial condition in the perturbed system. However, the decrease in overlap and committor value is not drastic indicating that small variations in the Hamiltonian will not disrupt the process of wavefunction optimization. Similarly, in Fig. \ref{fig6} an optimal wavefunction with yield of 98\% or higher could be identified for all perturbed systems, indicating that the chosen parameters in this work are not unique in providing opportunities for optimization. 

Perturbations of other parameters produced similar results or, in the case of $\omega_c$ and $E_2$, showed much greater stability of the optimized wavefunction to parameter perturbation. A sufficiently large perturbation of any parameter will, however, completely change the eigenstate ordering and, though an optimal coherent initialization may still exist, it will no longer involve a superposition of eigenstates 13 and 14 and direct comparisons cannot be drawn.

\subsection*{Appendix C: Derivation of Quantum Committor}
In this section, we will derive Eq. (\ref{eq:committor}) for the committor from the main text.
This derivation will proceed by first defining the quantum jump unraveling corresponding to the partial secular master equation, Eq. (\ref{eq:partial_secular}), and using this stochastic Poisson process to define a commitor equivalently to standard classical Poisson processes.
Then, we define an auxiliary stochastic process that implements absorbing boundary conditions in the reactant and product non-secular blocks while leaving dynamics unchanged outside of this manifold, which conveniently expresses the commitor.
Finally, we average over the realizations of the jump process to obtain a new quantum master equation with absorbing boundary conditions yielding Eq. (\ref{eq:committor_component}). 
Crucially, we see that this committor, and the resulting interpretation, are independent of the unraveling used in the derivation.

Equation (\ref{eq:partial_secular}) is a quantum master equation that we can equivalently solve as the average over a stochastic Poisson process for a density matrix, $\hat{\rho}$, called the quantum jump unraveling,
\begin{subequations}
    \label{eqs:rho_SDE}
    \begin{eqnarray}
        \label{eq:rho_SDE}
        d\hat{\rho}(t) &=& \left[\L_0 + \sum_\Omega \gamma_\Omega\left( \A_\Omega +\langle n_\Omega(\hat{\rho})\rangle\right) \right]\hat{\rho} dt \notag \\
        &&+ \sum_\Omega \left(\frac{\J_\Omega}{\langle n_\Omega(\hat{\rho})\rangle} 
        - \Id \right)\hat{\rho} dN_\Omega(\hat{\rho})
    \end{eqnarray}
    \begin{equation}
        \label{eq:N_Expect}
        \Ex \left[dN_\Omega(\hat{\rho})\right] =\gamma_\Omega \langle n_\Omega (\hat{\rho})\rangle dt
    \end{equation}
    \begin{equation}
        \label{eq:n_mean}
        \langle n_\Omega(\hat{\rho}) \rangle \equiv \Tr\{\J_\Omega \hat{\rho}\} = -\Tr\{\A_\Omega \hat{\rho}\},
    \end{equation}
\end{subequations}
where the double index $\Omega=(k,\overline{\omega})$ has been introduced for brevity, $\gamma_\Omega \equiv \gamma_k(\bar{\omega}, \bar{\omega})$, $\Id$ is the identity superoperator, $\L_0 = -i[H_S +H_{LS}, \cdot]$ is the unitary system propagator, $\J_\Omega = S_k(\overline{\omega}) \cdot S_k^*(\overline{\omega})$ is the jump operator describing the change in system state due to observing a change in bath eigenstate, and $\A_\Omega = \lbrace{S_k^*(\overline{\omega}) S_k(\overline{\omega}) , \cdot \rbrace}$ is the anti-unitary (sometimes called Zeno) drift due to not observing a change in bath eigenstate.
The differentials $dN_\Omega$ are the increments of independent Poisson processes with state dependent rates given by Eq. (\ref{eq:N_Expect}).
Each realization of the stochastic process $\hat{\rho}$ corresponds to a different history of bath measurement outcomes.
In the partial secular limit we know that the jumps couple superpositions in one secular manifold to superpositions in another, decoupled from intra-manifold coherences.
We will use the decorator $\hat{\cdot}$ throughout to indicate random variables defined for single realizations of the unraveling.
This stochastic differential equation is connected to Eq. (\ref{eq:partial_secular}) by $\Ex[\hat{\rho}]=\rho$, where $\Ex$ denotes an average over realizations of the stochastic unraveling.
Conditional density matrices remain normalized but the evolution is nonlinear in $\sigma$ since $\langle n_\Omega(\hat{\rho})\rangle$ is itself a linear function of $\hat{\rho}$.

We now proceed to define the committor equivalently to the case of a classical jump process.
We are able to use this approach since the quantum jump unraveling is simply a classical jump process over an unusual state-space of density matrices.
First, we select a reactant non-secular block $A$ and a product block $B$.
The committor can be constructed from the trajectory observable $\hat{P}_{A|B}[\hat{\rho}(t)]$ which is $1$ if the trajectory visits $B$ before $A$ and $0$ otherwise.
If we average over all trajectories with initial state $\bm{\rho}$, we obtain the committor $P_{A|B}(\bm{\rho})=\Ex\left[\hat{P}_{A|B}[\hat{\rho}(t)]|\hat{\rho}(0)=\bm{\rho}\right]$.
The commutator can then be computed by sampling realizations of this stochastic process starting in state $\bm{\rho}$ and computing the average over the trajectory ensemble.
However, since the stochastic equation is nonlinear in $\hat{\rho}$ this must be recomputed for every initial state $\bm{\rho}$ repeatedly which can rapidly become costly since the size of the state space grows rapidly with increasing system dimensions.

It is therefore desirable to obtain an expression for the committor using a quantum master equation which is linear in $\bm{\rho}$.
We could then compute the contribution to the committor from each density matrix element $V_{B|A}(\rho_{ij})$ separately.
The linearity of the master equation would then allow us to compute the commutator for any $\bm{\rho}=\sum_{ij}c_{ij}\rho_{ij}$, a superposition of the density matrix elements $\rho_{ij}$, by evaluating the same superposition of  $V_{B|A}(\rho_{ij})$.
To accomplish this, we define an auxiliary unraveling $\hat{\tilde{\rho}}$, identical to Eq. (\ref{eqs:rho_SDE}), but removing all jumps out of blocks $A$ and $B$.
This stochastic process has the same dynamics as the original unraveling outside of $A\cup B$ but does not allow population to leave $A$ or $B$ once it enters either block.
If we ignore all dynamics within $A$ and $B$ and consider only the total population of the blocks, this procedure implements two absorbing boundary conditions at the reactant and product states.

Using the absorbing auxiliary unraveling, $\hat{\tilde{\rho}}$, we can easily rewrite the trajectory observable,
\begin{equation}
    \label{eq:single_traj}
    \hat{P}_{A|B}[\hat{\rho}(t)] = \hat{P}_{A|B}[\hat{\tilde{\rho}}(t)] = \lim_{t\to\infty}\Tr\{\P_B\hat{\tilde{\rho}}(t)\}
\end{equation}
as the steady-state population of non-secular product block $B$, written in terms of the $B$ projection operator $\P_B = \sum_{b\in B} \ket{b}\bra{b}$.
For each trajectory, this quantity is always exactly $1$ or $0$ since the probability of simultaneous jumps vanishes, and in the non-secular master equation each jump can only connect one pair of blocks.
Therefore, it is impossible for the unraveling to simultaneously enter blocks $A$ and $B$.

The introduction of the absorbing process has now reduced the calculation of the committor to the expectation of a steady-state population.
This can be equivalently done {\it after} averaging over the stochastic processes to obtain an auxiliary quantum master equation with generator $\L'$, yielding Eq. (\ref{eq:committor}) in the main text.
Notably, this definition of the committor is independent of the choice of unraveling used in the derivation, indicating that it captures the properties of the underlying dynamical process independent of any assumed measurement protocol on the bath.

\section*{References}
\bibliography{ref}

\end{document}